\documentclass[sigconf]{acmart}
\makeatletter
\let\@authorsaddresses\@empty
\makeatother
\renewcommand\footnotetextcopyrightpermission[1]{}
\usepackage{caption,subcaption}
\usepackage[ruled,vlined]{algorithm2e}
\usepackage{listings}
\usepackage{listings-rust}
\usepackage[utf8]{inputenc}
\usepackage[english]{babel}
\usepackage{amsthm}
\usepackage{multirow}
\theoremstyle{definition}
\newtheorem{definition}{Definition}[section]
\usepackage{enumerate}
\usepackage{graphicx}
\usepackage{color}
\usepackage{colortbl}

\definecolor{mygray}{gray}{.9}

\acmConference[arxiv]{arxiv pre-print versions}{2021}{}

\author{Mohan Cui}
\author{Chengjun Chen\textsuperscript{*}}
\thanks{\textsuperscript{*}Chengjun Chen and Mohan Cui have equal contributions to this work.}
\affiliation{School of Computer Science,\\ Fudan University}

\author{Hui Xu\textsuperscript{+}}\thanks{\textsuperscript{+}corresponding author}
\author{Yangfan Zhou}
\affiliation{School of Computer Science,\\ Fudan University}

\begin{document}
\title{SafeDrop: Detecting Memory Deallocation Bugs of Rust Programs via Static Data-Flow Analysis}

\begin{abstract}
Rust is an emerging programming language that aims to prevent memory-safety bugs. However, the current design of Rust also brings side effects which may increase the risk of memory-safety issues. In particular, it employs OBRM (ownership-based resource management) and enforces automatic deallocation of unused resources without the garbage collector. It may therefore falsely deallocate reclaimed memory and lead to use-after-free or double-free issues. In this paper, we study the problem of invalid memory deallocation and propose \textit{SafeDrop}, a static path-sensitive data-flow analysis approach to detect such bugs. Our approach analyzes each API of a Rust crate iteratively by traversing the control-flow graph and extracting all aliases of each data-flow. To guarantee precision and scalability, we leverage a modified Tarjan algorithm to achieve scalable path-sensitive analysis, and a cache-based strategy to achieve efficient inter-procedural analysis. Our experiment results show that our approach can successfully detect all existing CVEs of such issues with a limited number of false positives. The analysis overhead ranges from 1.0\% to 110.7\% in comparison with the original compilation time. We further apply our tool to several real-world Rust crates and find 8 Rust crates involved with invalid memory deallocation issues.
\end{abstract}

\maketitle

\section{Introduction}
Rust is an emerging programming language with attractive features for memory-safety protection yet providing comparative efficiency as C/C++. It achieves the goal by dividing Rust programming language into safe Rust and unsafe Rust. The boundary is an `unsafe' marker that confines the risks of memory-safety issues (\textit{e.g.,} dereferencing raw pointers) within unsafe code only~\cite{DieHard}. In a narrow sense, Rust guarantees the soundness of safe Rust with no risks of introducing memory-safety issues~\cite{SystemRust}. Due to these advantages, many real-world projects start to embrace Rust, such as Servo~\cite{anderson2016engineering} and TockOS~\cite{levy2017tock}. Although the feedback of Rust's effectiveness in preventing memory-safety bugs is positive, there still exist large amounts of such bugs in real-world projects~\cite{qin2020understanding,Evans2020is,xu2020memory}.

This work studies a specific type of memory-safety bugs found in real-world Rust crates (projects) related to RAII (resource acquisition is initialization). In particular, Rust employs a novel OBRM (ownership-based resource management) model, which assumes the resource should be allocated when creating an owner and deallocated once its owner goes out of the valid scope. Ideally, this model should be able to prevent dangling pointers~\cite{undangling} and memory leakages~\cite{UAFDP}, even when a program encounters exceptions. However, we observe that many critical bugs of real-world Rust crates are associated with such automatic deallocation scheme, \textit{e.g.,} it may falsely drop some buffers that are still being used and incur use-after-free bugs (\textit{e.g.,} CVE-2019-16140), or may falsely drop dangling pointers and cause double free (\textit{e.g.,} CVE-2019-16144).

In general, memory deallocation bugs are triggered by unsafe code. Unsafe APIs are necessary in Rust to provide the low-level control and abstraction over implementation details~\cite{Evans2020is}. However, misusing unsafe APIs can invalidate the soundness of ownership-based resource management system and may cause undefined behaviors. For example, an unsafe API may lead to memory reclaim of shared aliases and dropping one instance would incur dangling pointers for the remaining aliases~\cite{StackedBorrows}. Moreover, the interior unsafe~\cite{qin2020understanding} in Rust, allowing a function that has unsafe code only in internal and can be called as safe function, may have potential memory-safety issues inside. The current Rust compiler has done little regarding the memory-safety risks of unsafe code but simply assumes developers should be responsible for employing them. As memory-safety is the most important feature promoted by Rust, reducing such risks is exceedingly important if possible.

To tackle the problem, this paper proposes SafeDrop, a static path-sensitive data-flow analysis approach for detecting memory-safety bugs due to the automatic deallocation mechanism. SafeDrop analyzes each API of a Rust crate iteratively and scrutinizes whether each drop statement in Rust MIR (mid-level intermediate representation) is safe to launch. Since the underlying problem of alias analysis is NP-hard if it could be decidable~\cite{landi}, we have adopted several designs to improve the scalability of our approach while not sacrificing much precision. Our approach adopts a meet-over-paths (MOP)~\cite{MOP} method and extracts all valuable paths of each function based on a modified Tarjan algorithm~\cite{10.1145/800061.808753}, which is effective to eliminate redundant paths in cycles with identical alias relationships. For each path, we extract the sets of all aliases in a flow-sensitive manner and analyze the safety of each drop statement accordingly. When encountering function calls, we recursively perform SafeDrop on the callee, and analyze the alias relations between arguments and return value. To avoid duplicated analysis, we cache and reuse the obtained aliasing result of each function.

We have implemented our approach as a query (pass) of Rust compiler v1.52 and conducted real-world experiments on existing CVEs of such types. Experimental results show that SafeDrop can successfully recall all these bugs with a limited number of false positives. The analysis overhead ranges from 1.0\% to 110.7\% in comparison with the original compilation time. We further apply SafeDrop to several other Rust crates and find 8 crates that have invalid memory deallocation issues previously unknown.

We summarize the contribution of this paper as follows.

\begin{itemize}
\item Our paper serves as the first attempt to study memory deallocation bugs of Rust programs related to the side effect of RAII. We systematically discuss the problem and extract several common patterns of such bugs. Although our work focuses on Rust, the problem may also exist in other programming languages with RAII.

\item We have designed and implemented a novel path-sensitive data-flow analysis approach for detecting memory deallocation bugs. It employs several careful designs in order to be scalable while not sacrificing much precision, including a modified Tarjan algorithm for path-sensitive analysis and a cache-based strategy for inter-procedural analysis. 

\item We have conducted real-world experiments with existing Rust CVEs and verified the effectiveness and efficiency of our approach. Moreover, we find 8 Rust crates involved with invalid memory deallocation issues previously unknown.
\end{itemize}

\section{Preliminaries}
This section presents the background of the problem, including the memory management model of Rust, the basics of Rust compiler and the borrow checker of Rust as well as its limitations.

\subsection{Memory Management of Rust}
Rust introduces a novel ownership-based resource management system to manage memory, and this model assumes each variable has exclusive ownership for its allocated memory. Ownership can be borrowed as references in either mutable or immutable manner with several restrictions (lifetime rules). It requires that reference cannot outlive its referent and the mutable reference cannot be aliased. Rust also provides the traditional raw pointers alike references without previous requirements. However, any operations of raw pointers that may violate the memory-safety are deemed as unsafe and need to be wrapped with the `unsafe' marker \textit{e.g.,} dereferencing raw pointers.

In type system, Rust divides types into two mutex kind of trait: \texttt{Copy} trait for stack-only data and \texttt{Drop} trait~\cite{klabnik2019rust} for others. These traits are important to manage memory and determine how Rust processing rvalues to generate lvalues in assignments, as well as parameter passing and value returning. If the rvalue has \texttt{Copy} trait, Rust will duplicate (copy) it in the stack and the older variable is still usable. Otherwise, it will transfer (move) the ownership from rvalue to lvalue and the older variable is no longer available.

The memory management idea of Rust shares many similarities with the intelligent pointers of C++, and it benefits Rust in realizing automatic RAII, \textit{i.e.,} resources are bounded with valid scopes~\cite{ramananandro2012mechanized}. In particular, each drop-trait variable or temporary is associated with a fixed drop scope\footnote{https://doc.rust-lang.org/reference/destructors.html}. When an initialized variable or temporary goes out of the drop scope, the destructor will recursively drop its fields in order.

Because Rust intentionally manages the ownership of each drop-trait instance at each program point, it can free unused resources without garbage collection by automatically running their destructors once control-flow leaves the drop scope.

\subsection{Basics of Rust Compiler}

\begin{figure}[t!b]
\includegraphics[width=0.46\textwidth]{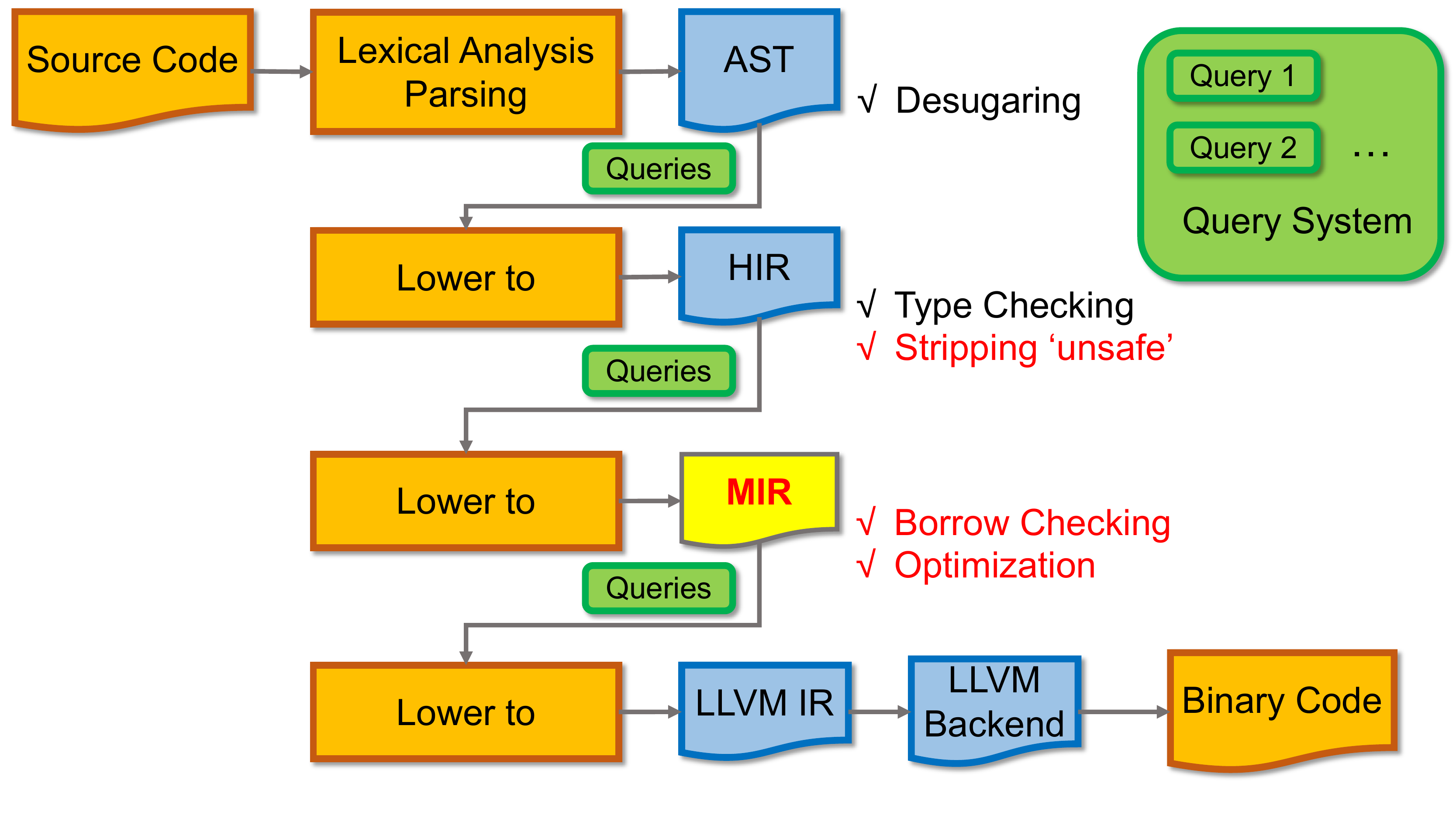}
\caption{Procedures of the Rust compiler.}
\label{fig:compiler}
\end{figure}

Rust compiler defines a query system to perform demand-driven compilation, which is different from traditional pass-based systems~\cite{stallman2002gnu}. Figure~\ref{fig:compiler} presents several main steps for compiling Rust programs, and the key design for detecting memory-safety violations is based on MIR. The syntax of MIR is in SSA (single static assignment) form, which is similar to the traditional LLVM IR (intermediate representation)~\cite{lattner2008llvm}. For simplicity, Listing~\ref{list:mir_syntax} presents the unique syntax of Rust MIR that lays the foundation for achieving ownership-based resource management, and the full syntax system is available in rfcs\#1211\footnote{https://rust-lang.github.io/rfcs/1211-mir.html}.

According to the syntax, a basic block is composed of several statements and a single terminator. The statements mainly consist of three forms: assignment, storage-live statement, and storage-dead statement. The assignment assigns the value of \texttt{RValue} to \texttt{LValue} in different ways. In particular, \texttt{move LValue} means transferring the ownership of \texttt{LValue} to \texttt{RValue}; \texttt{\& mut LValue} means borrowing \texttt{LValue} as a mutable reference; and \texttt{* LValue} means creating an immutable raw pointer towards \texttt{LValue}. The storage-live statement and storage-dead statement respectively represent the start and the end of a live range for the storage of a variable. The terminator of each basic block describes how it connects to subsequent blocks. \texttt{Drop(Value, BB0, BB1)} means invoking the destructor of \texttt{Value} once \texttt{Value} goes out of its drop scope. To achieve automatic resource deallocation when encountering exceptions, the terminators that may incur exceptions are common in Rust MIR, \textit{i.e.,} an extra exception handling block \texttt{BB1} is added for taking over the stack unwinding process. In particular, \texttt{SwitchInt()} is a special terminator that we will discuss in Section 4.

\lstset{language=Rust, style=colouredRust}
\begin{figure}[t]
\begin{lstlisting}[language=Haskell, caption = Core syntax of Rust MIR. \label{list:mir_syntax}] 
BasicBlock := {Statement} Terminator
Statement := LValue = RValue | StorageLive(Value)
	| StorageDead(Value) | ...
LValue := LValue | LValue.f | *LValue | ...
RValue := LValue | move LValue 
	| & LValue | & mut LValue
	| * LValue | * mut LValue
	| ... 
Terminator := Goto(BB) | Panic(BB)
	| Return | Resume | Abort
	| If(Value, BB0, BB1)
	| LVALUE = (FnCall, BB0, BB1)
	| Drop(Value, BB0, BB1)
	| SwitchInt(Value, BB0, BB1, BB2, ...)
	| ...        
\end{lstlisting}
\end{figure}

\subsection{Borrow Checker of Rust}
Rust MIR introduces a borrow (reference) checker based on lifetime inference. Non-lexical lifetime (NLL) can save much programmers' effort for tediously specifying the lifetime in a fine-grained manner and it determines whether a reference is valid to use. The mechanism is officially documented in rfcs\#2094\footnote{https://rust-lang.github.io/rfcs/2094-nll.html}, and we highlight the key idea as below. NLL extracts a set of constraints based on the considerations of liveness, subtyping, and reborrowing requirements. \textit{Liveness constraint} means a reference should be valid from declaration to last use; \textit{subtyping constraint} means the lifetime of a reference should not exceed its referent; \textit{reborrow constraint} means the lifetime of a reference, that is created from the instance by dereferencing other references, should not exceed its referent. Rust compiler then solves the constraints established by NLL and computes the minimal lifetime of each reference via fixed-point iteration. In this way, it can generate an optimized solution. Thus, the borrow checker can detect memory-safety violations that are infringing the lifetime rules previously made for references, \textit{i.e.,} the reference outlives its referent. Based on these information, the safety of references can be guaranteed in Rust compiler.

\subsection{Soundness Limitations of Rust Compiler}
Now, we introduce the memory management model and the borrow checker in Rust. As shown in Figure~\ref{fig:compiler}, Rust compiler strips `unsafe' markers before lowering the code to MIR, thus the borrow checker is valid for both safe Rust and unsafe Rust. However, this system exists several drawbacks. The sound lifetime inference and borrow checker ensure the safety of references only. Any unsafe code interacted with raw pointers can breach the safety promise and may lead to memory reclaim. Moreover, the memory management model cannot discriminate the alias relationship between multiple drop-trait instances, because it only associates every instance with a fixed drop scope. Since there is no such algorithm for ensuring the safety of raw pointers and memory reclaim, this is the primary reason that leads to the memory deallocation bugs studied in this paper.

\section{Problem Statement}\label{sec:problem}

\subsection{Motivating Example}

\lstset{language=Rust, style=scriptsize}
\begin{figure*}[t]
\centering
\begin{subfigure}[b]{0.4\textwidth}
\begin{lstlisting}[language = Rust,] 
fn genvec() -> Vec<u8> {
    let mut s = String::from("a tmp string");
    let ptr = s.as_mut_ptr();
    let v;
    unsafe{
        v = Vec::from_raw_parts(
        	ptr, s.len(), s.len());
    }
    // mem::forget(s); // do not drop s
    // otherwise, s is dropped before return 
    return v;
}

fn main() {
    let v = genvec();
    // use v -> use after free
    // drop v before return -> double free
}
\end{lstlisting}
\caption{Source code of dropping aliases.}
\label{fig:src_baddrop}
\end{subfigure}
\begin{subfigure}[b]{0.59\textwidth}
	\includegraphics[width=\textwidth]{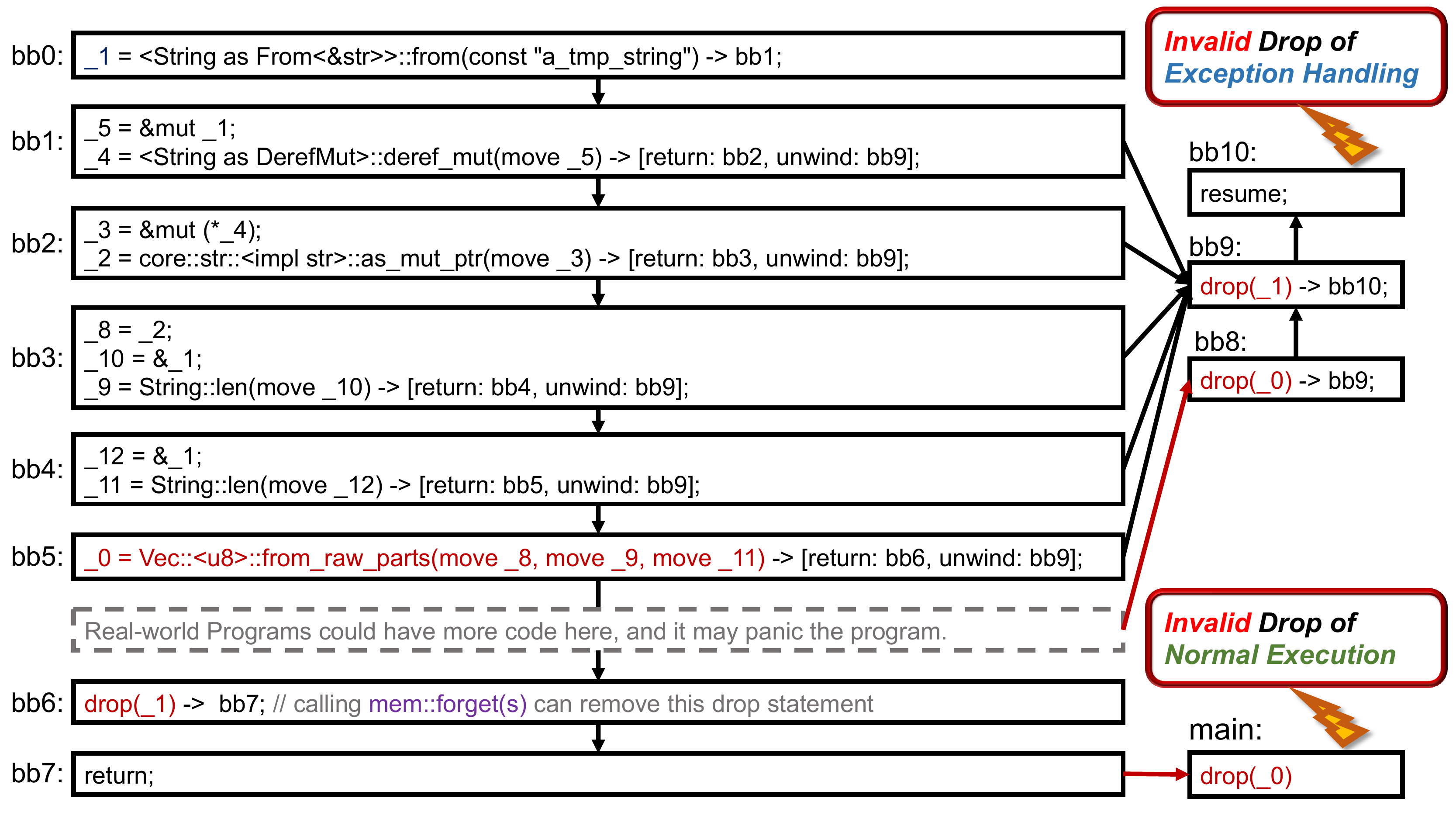}
	\caption{MIR form and CFG of Figure~\ref{fig:src_baddrop}.}
    \label{fig:mir_baddrop}
\end{subfigure}

\begin{subfigure}[b]{0.4\textwidth}
\begin{lstlisting}[language = Rust, ] 
struct Foo{
    vec : Vec<i32>,
}

impl Foo {
    pub unsafe fn read_from(src: &mut Read) -> Foo{
    	let mut foo = mem::uninitialized::<Foo>();
    	let s = slice::from_raw_parts_mut(
    		&mut foo as *mut _ as *mut u8, 
    		mem::size_of::<Foo>());
    	src.read_exact(s);
        foo
    }
}
\end{lstlisting}
\vspace{1cm}
\caption{Source code of dropping uninitialized memory.}
\label{fig:src_baddrop2}
\end{subfigure}
\begin{subfigure}[b]{0.59\textwidth}
	\includegraphics[width=\textwidth]{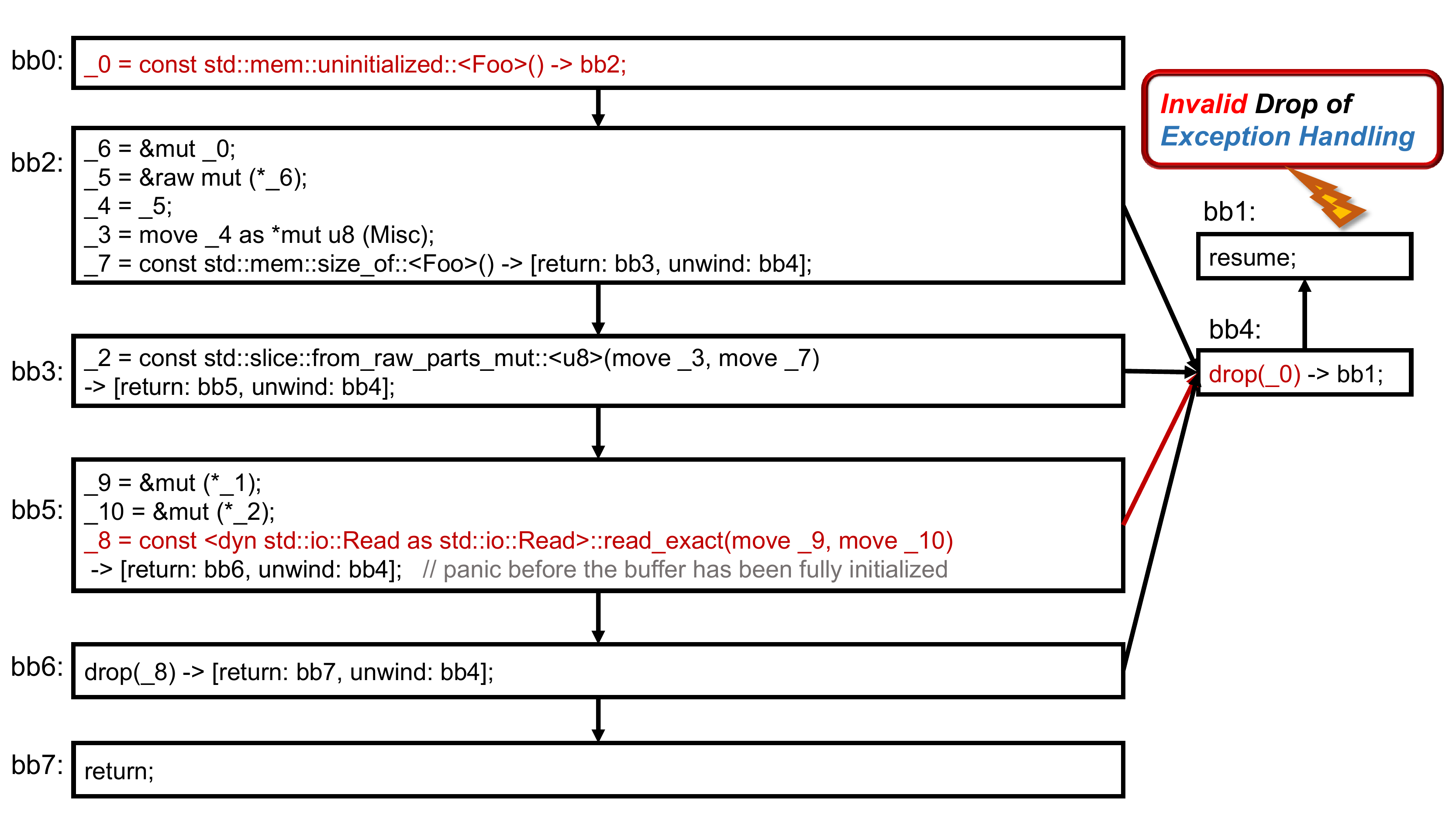}
	\caption{MIR form and CFG of Figure~\ref{fig:src_baddrop2}.}
    \label{fig:mir_baddrop2}
\end{subfigure}
\caption{Motivating example of memory-safety issues incurred by automatic deallocation.}
\label{fig:baddrop}
\end{figure*}

Rust enforces RAII and releases unused resources automatically. In practice, this mechanism may falsely drop some buffers and is prone to memory-safety issues. Based on the occurrence of a buggy \texttt{Drop()} terminator, we categorize such problems into two situations: invalid drop in normal execution path and invalid drop in exception handling path.

\subsubsection{\textbf{Invalid Drop of Normal Execution}}
Such a bug happens if the culpriting \texttt{Drop()} terminator locates in a normal execution path, and the parameter of \texttt{Drop()} is not safe to launch. We use Figure~\ref{fig:src_baddrop} as a proof-of-concept (PoC) example to demonstrate such cases.

The \texttt{genvec()} function is an interior unsafe constructor that creates vector \texttt{v} based on a raw pointer \texttt{ptr}. In this PoC, the string \texttt{s} and the vector \texttt{v} are sharing the same memory space. After the automatic deallocation of the string, the vector will contain a dangling pointer pointing to the released buffer, and using this vector later in the main function will incur use-after-free. As long as the vector is no longer used and the control-flow goes out of its drop scope, it will be dropped automatically and incur double free. There are many such bugs found in practice, such as CVE-2019-16140, which is use-after-free, and CVE-2018-20996 and CVE-2019-16144, which are double free.

Figure~\ref{fig:mir_baddrop} shows the equivalent description in MIR syntax. \texttt{\_1} in \texttt{bb0} creates a new string and the returned vector \texttt{\_0} in \texttt{bb5} is created based on \texttt{\_1} with an alias propagation chain \texttt{1->5->4->3->2->8->0}. Therefore, \texttt{\_0} contains an alias pointer of \texttt{\_1}. Namely, dropping \texttt{\_1} incurs dangling pointers of \texttt{\_0} and dropping \texttt{\_0} later in the main function incurs double free.

Why cannot borrow checker add the support for raw pointers to detect such issues? The problem lies in the trade-off design for dealing with function calls. For example, \texttt{\_2} is created by calling \texttt{deref\_mut(move \_5)}, and its parameter \texttt{\_5} is a mutable alias of \texttt{\_1}. For simplicity, Rust assumes that each parameter should either transfer its ownership to the callee for the drop-trait variable or duplicate a deep copy for the copy-trait variable, and the return value would no longer share the ownership with the alive variables remained. Such assumptions save the borrow checker from the complex inter-procedural analysis. As a trade-off, it leaves the holes for raw pointers in unsafe Rust. Note that \texttt{from\_raw\_parts()} is an unsafe function, and it is the culprit that leads to the memory reclaim in this example. However, the `unsafe' marker has been stripped away after lowering the code to MIR, and the Rust compiler cannot differentiate the safety boundary anymore. Moreover, the current Rust compiler does not add the alias checking support for raw pointers and not to mention a more complex data structure. Therefore, the problem cannot be fixed easily in the current framework.

\subsubsection{\textbf{Invalid Drop of Exception Handling}}
In some real-world bugs, memory deallocation issues only exist in the exception handling path, as the program panics and enters into the unwinding process. For example, CVE-2019-16880 and CVE-2019-16881 have double-free issues if the program panics; CVE-2019-15552 and CVE-2019-15553 may drop uninitialized memory in exception handling. Figure~\ref{fig:baddrop} presents this PoC to demonstrate the problem.

Suppose developers have fixed the bug by adding \texttt{mem::forget()} to the code in Figure~\ref{fig:src_baddrop} which prevents \texttt{drop(\_1)} in \texttt{bb6} in Figure~\ref{fig:mir_baddrop}. They may also add more statements (\textit{e.g.,} retrieving the content of the vector) between creating \texttt{v} and calling \texttt{mem::forget()}. In this way, these additional statements are still vulnerable program points. If the program panics in these points, according to the principle of RAII, Rust should deallocate resources during stack unwinding by continuously calling \texttt{Drop()}. Due to the existance of aliasing drop-trait instances, dropping these variables would lead to double free issues. Moreover, \texttt{mem::forget(s)} takes the entire ownership of the string \texttt{s}, it is generally used as late as possible so that other statements can still use the string.

Likewise, dropping uninitialized memory is also a popular issue during exception handling. Figure~\ref{fig:src_baddrop2} and \ref{fig:mir_baddrop2} demonstrate a PoC that applies an uninitialized buffer first and initializes it afterwards. However, if the program panics before the buffer has been fully initialized, the stack unwinding process would drop those uninitialized memory, which is similar to use-after-free if the buffer has pointers.

\subsection{Problem Definition}
We formalize the problem as follows. Supposing a programming language supporting RAII would automatically deallocate unused memory based on some static strategies. Due to the limitations of static analysis, such deallocations may cause memory-safety bugs. In particular, there are two types of invalid memory deallocations.

\begin{definition}[Dropping buffers in use]
If the algorithm falsely deallocates some buffers that will be accessed later, it would incur dangling pointers that are vulnerable to memory-safety issues, including use-after-free and double-free.
\end{definition}

\begin{definition}[Dropping invalid pointers]
If the invalid pointer is dangling, dropping the pointer would incur double free; if the invalid pointer points to an uninitialized memory containing pointer types, dropping the pointer may drop its nested pointers recursively and incur invalid memory access.
\end{definition}

The problem of invalid drop should be a common issue for programming languages that enforce RAII, such as C++ and Rust. However, it is more severe in Rust due to two reasons. Firstly, Rust emphasizes much more on memory safety, and such security issues are less tolerable. Secondly, since Rust has no garbage collector, it is very aggressive in resource recycling and memory-leakage prevention.

\subsection{Typical Patterns}

\lstset{language=Rust,}
\begin{figure}[t]
\begin{lstlisting}[language=C, caption = Typical patterns checked for invalid memory deallocation in SafeDrop. (UAF: use after free; DF: double free; IMA: invalid memory access)  \label{list:pattern}]
Pattern 1: GetPtr() -> UnsafeConstruct() -> Drop() -> Use()  => UAF
Pattern 2: GetPtr() -> UnsafeConstruct() -> Drop() -> Drop() => DF
Pattern 3: GetPtr() -> Drop() -> UnsafeConstruct() -> Use()  => UAF
Pattern 4: GetPtr() -> Drop() -> UnsafeConstruct() -> Drop() => DF
Pattern 5: GetPtr() -> Drop() -> Use() => UAF
Pattern 6: Uninitialized() -> Use()  => IMA
Pattern 7: Uninitialized() -> Drop() => IMA
\end{lstlisting}
\end{figure}

We illustrate several typical patterns of such bugs. Note that since the data-flow approaches for normal execution paths and exception handling paths are similar, we do not further differentiate them in these patterns. Listing~\ref{list:pattern} summarizes 7 typical patterns of invalid memory deallocation from the view of statement sequences. Pattern 1 and pattern 2 corresponds to our PoC of Figure~\ref{fig:mir_baddrop} that incurs use-after-free (UAF) or double-free (DF). We employ \texttt{GetPtr()} as a general representation for obtaining a raw pointer to a drop-trait instance, and employ \texttt{UnsafeConstruct()} to denote unsafe function calls for constructing new aliasing instances (\textit{e.g.,} \texttt{from\_raw\_part()}). Pattern 3 and pattern 4 are similar but switch the order of \texttt{UnsafeConstruct()} and \texttt{Drop()}. Pattern 5 has no \texttt{UnsafeConstruct()} but uses the dangling pointer directly and incurs use-after-free. Pattern 6 and pattern 7 correspond to our PoC of Figure~\ref{fig:mir_baddrop2} that relates to uninitialized memory. We employ \texttt{Uninitialized()} to represent the constructor of drop-trait instance without initialization. Either using uninitialized memory or dropping it directly would be vulnerable to invalid memory access (IMA).

\section{Approach}
In this section, we describe our approach for detecting invalid memory deallocation problems described in Section~\ref{sec:problem}.

\subsection{Overall Framework}

\begin{figure}[t!b]
\includegraphics[width=0.45\textwidth]{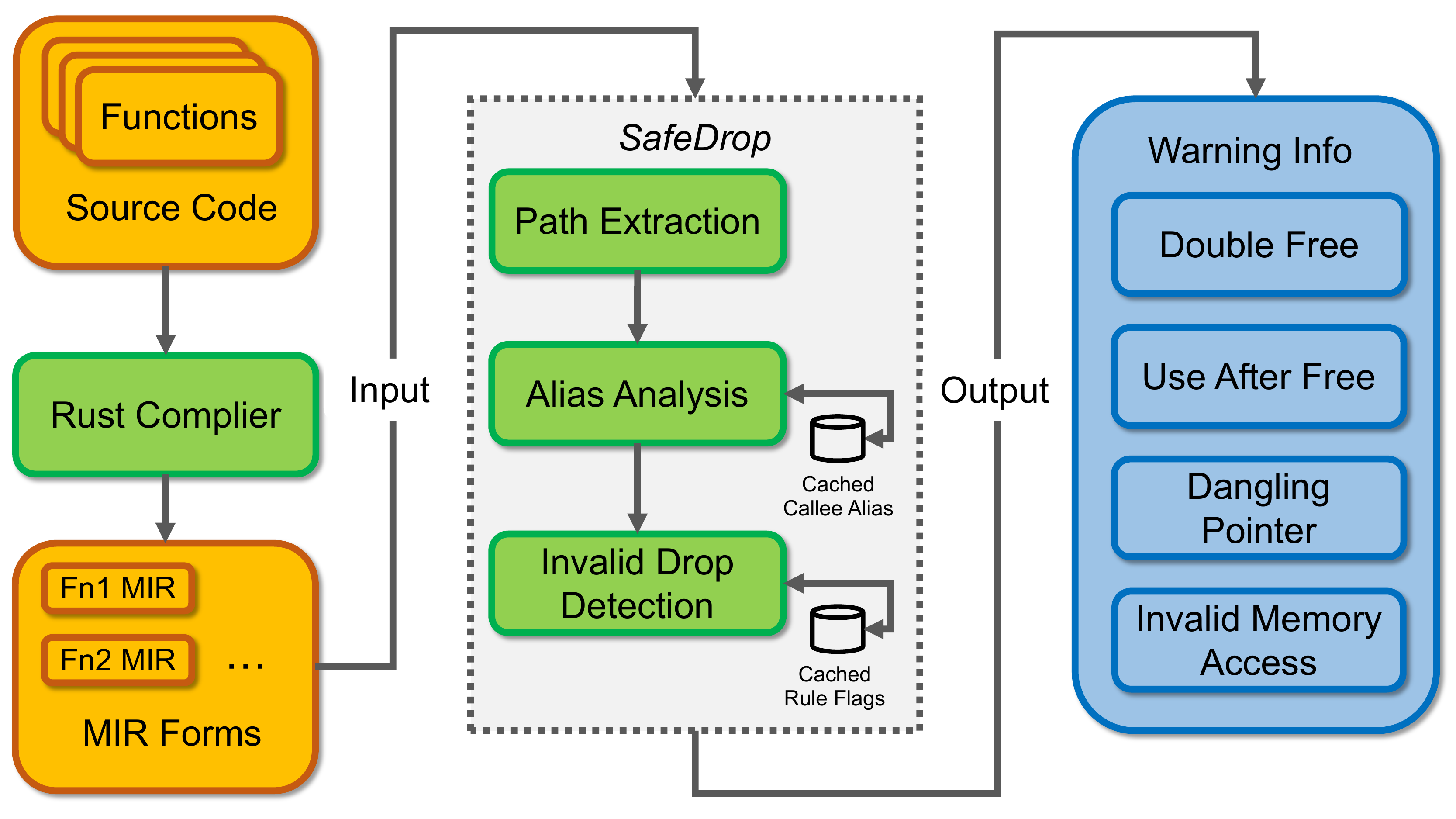}
\caption{Overall framework of SafeDrop.}
\label{fig:framework}
\end{figure}

We tackle the problem with a path-sensitive data-flow analysis approach and integrate it into the compiler, named SafeDrop. Figure~\ref{fig:framework} overviews the framework of our approach. It inputs the MIR of each function and outputs warnings of potential invalid memory deallocation issues along with the buggy code snippets. We discuss the key steps of SafeDrop as follows.

\begin{itemize}
\item \textbf{Path Extraction:} SafeDrop adopts a meet-over-paths method to achieve path-sensitivity. Since the paths of a function could be infinite, we employ a novel method based on Tarjan algorithm to merge redundant paths and generate the spanning tree. This step traverses this tree and finally enumerates all valuable paths. 
\item \textbf{Alias Analysis:} SafeDrop is field-sensitive. This step analyzes the alias relationship among the variables and the fields of the composite types for each data-flow. SafeDrop is also inter-procedural with context-insensitivity. It caches and reuses the obtained alias relationships of the callees between return value and arguments.
\item \textbf{Invalid Drop Detection:} Based on the alias sets established before, this step searches vulnerable drop patterns for each data-flow and records the suspicious code snippets.
\end{itemize}

In the following subsections, we will expand these steps by explaining more details.

\subsection{Path Extraction}
SafeDrop adopts a meet-over-paths method to improve the precision rather than using the traditional semi-lattice scheme~\cite{kleene1954upper}. It traverses the control-flow graph of a function and enumerates all valuable paths. In SafeDrop, the concept of a valuable path has two criteria: 1) a valuable path should be a unique set of code blocks with a start node (function entrance) and an exit node, and 2) the set of a valuable path should not be the subset of any other valuable paths. If there are conflicting paths for the second criterion, we only select the path with more unique blocks. In this way, we do not have to traverse over cycles repeatedly but only need to consider the maximum set of cycled blocks (as a strongly connected component). The validity can be self-explained according to the alias analysis rules for the code of SSA form~\cite{Andersen94programanalysis}. 

\begin{algorithm}[t]
	\caption{Main Algorithm of SafeDrop.}
	\label{algo:path}
	\LinesNumbered
	\SetKwData{data}{data}
	\SetKwData{node}{node}
	\SetKwData{child}{child}
	\SetKwData{children}{children}
	\SetKwData{cfg}{cfg}
	\SetKwData{st}{st}
	\SetKwData{sccs}{sccs}
	\SetKwData{result}{res}
	\SetKwData{false}{false}
	\SetKwData{root}{root}

	\SetKwFunction{Traverse}{Traverse}
	\SetKwFunction{GetChildren}{GetChildren}
	\SetKwFunction{Tarjan}{Tarjan}
	\SetKwFunction{GenerateTree}{GenerateTree}
	\SetKwFunction{ResolveConflict}{ResolveConflict}
	\SetKwFunction{ContainsConflict}{ContainsConflict}
	\SetKwFunction{AnalyzeAlias}{AnalyzeAlias}
	\SetKwFunction{DetectUnsafeDrop}{DetectUnsafeDrop}
	\SetKwFunction{IsNone}{IsNone}
	\SetKwFunction{Merge}{Merge}
	\SetKwFunction{PrintWarning}{PrintWarning}
	\SetKwFunction{Cache}{Cache}
	
	\SetKwProg{Fn}{Function}{}{end}	
	
	\BlankLine
	\BlankLine
    
    \KwData{{\cfg: the control-flow graph of the analyzing function;} {\data: the struct of the entire storage data in SafeDrop including alias relationships, type filter and taint sets;} {\result: the analysis results of the invalid drop detection flags and the alias relationship between arguments and return value;}}
    
    \BlankLine
    \BlankLine
    
    \result $\leftarrow$ \false\;
	\sccs $\leftarrow$ \Tarjan{\cfg}\tcp*[r]{extract all strongly connected components based on Tarjan algorithm}
	\st $\leftarrow$ \GenerateTree{\cfg, \sccs}\tcp*[r]{derive spanning tree}
	\ForEach{\node in \st} { 
	    \If{\st.\ContainsConflict{\node}} {
	        \st $\leftarrow$ \ResolveConflict{\st, \node}; 
	    }
	}
	\Traverse{\data, \result, \st.\root}\;
	\PrintWarning{\result}\tcp*[r]{output the warnings}
	\Cache{\result}\tcp*[r]{cache the result}
	
	\BlankLine
	\tcc*[h]{recursive traversal over the spanning tree}\\
	\Fn{\Traverse{\data, \result, \node}}{
	    \AnalyzeAlias{\data, \node}\tcp*[r]{in section 4.3}
	    \DetectUnsafeDrop{\data, \node}\tcp*[r]{in section 4.4}
		\children $\leftarrow$ \GetChildren{\node}\;
		\If{\children.\IsNone{}}{
			\Merge{\result}\tcc*[r]{meet over paths}
			\Return
		}
		\ForEach{\child in \children}{
			\Traverse{\data, \result, \child}\tcp*[r]{backup - recover}
		}
	}
	
\end{algorithm}

\begin{figure*}[t]
\begin{subfigure}[t]{0.49\textwidth}
\centering
	\includegraphics[width=\textwidth]{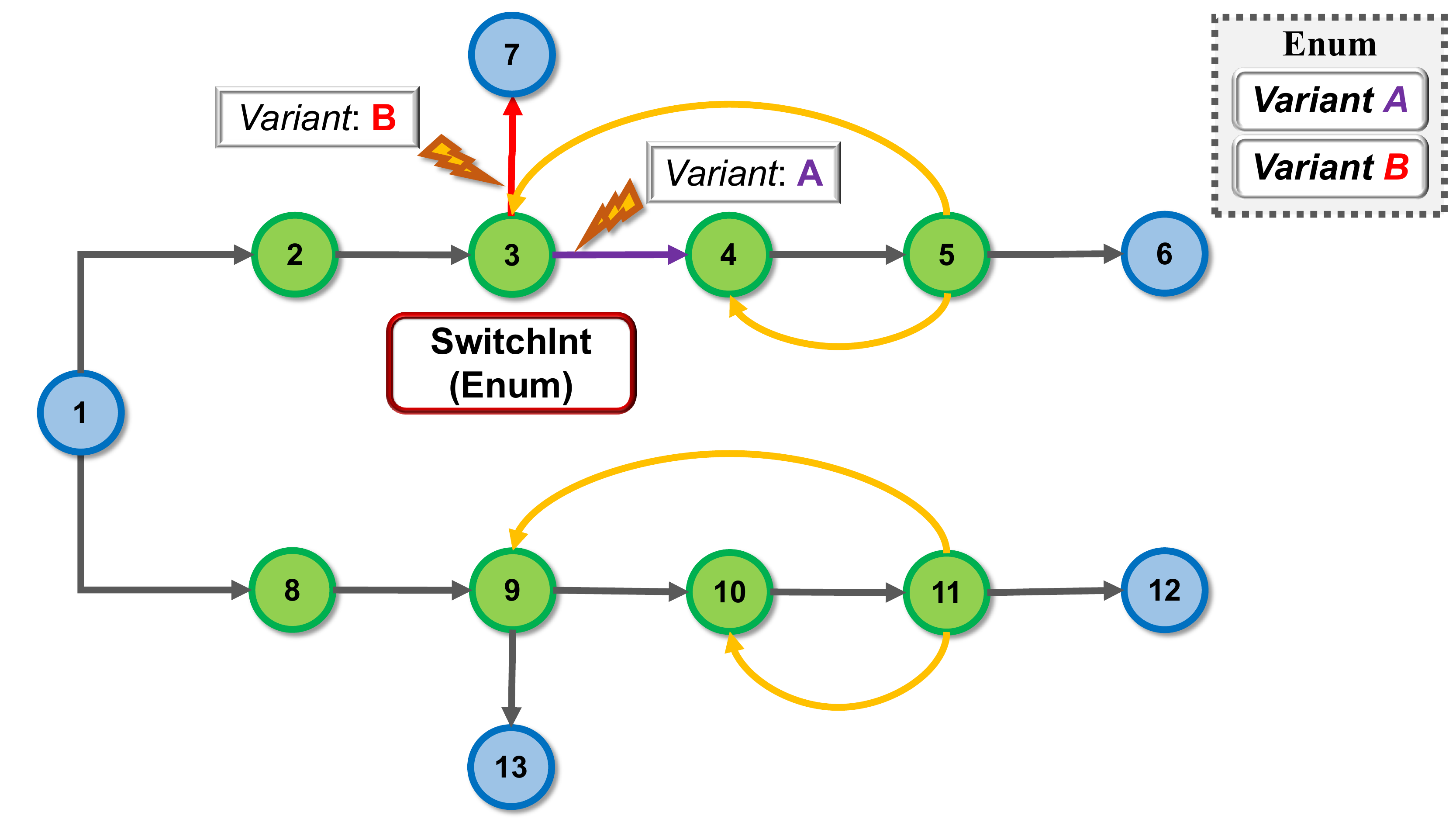}
	\caption{Control-flow graph example with SwitchInt() terminator.}
    \label{fig:control-flow-graph}
\label{fig:contro-flow-graph}
\end{subfigure}
\begin{subfigure}[t]{0.49\textwidth}
\centering
	\includegraphics[width=\textwidth]{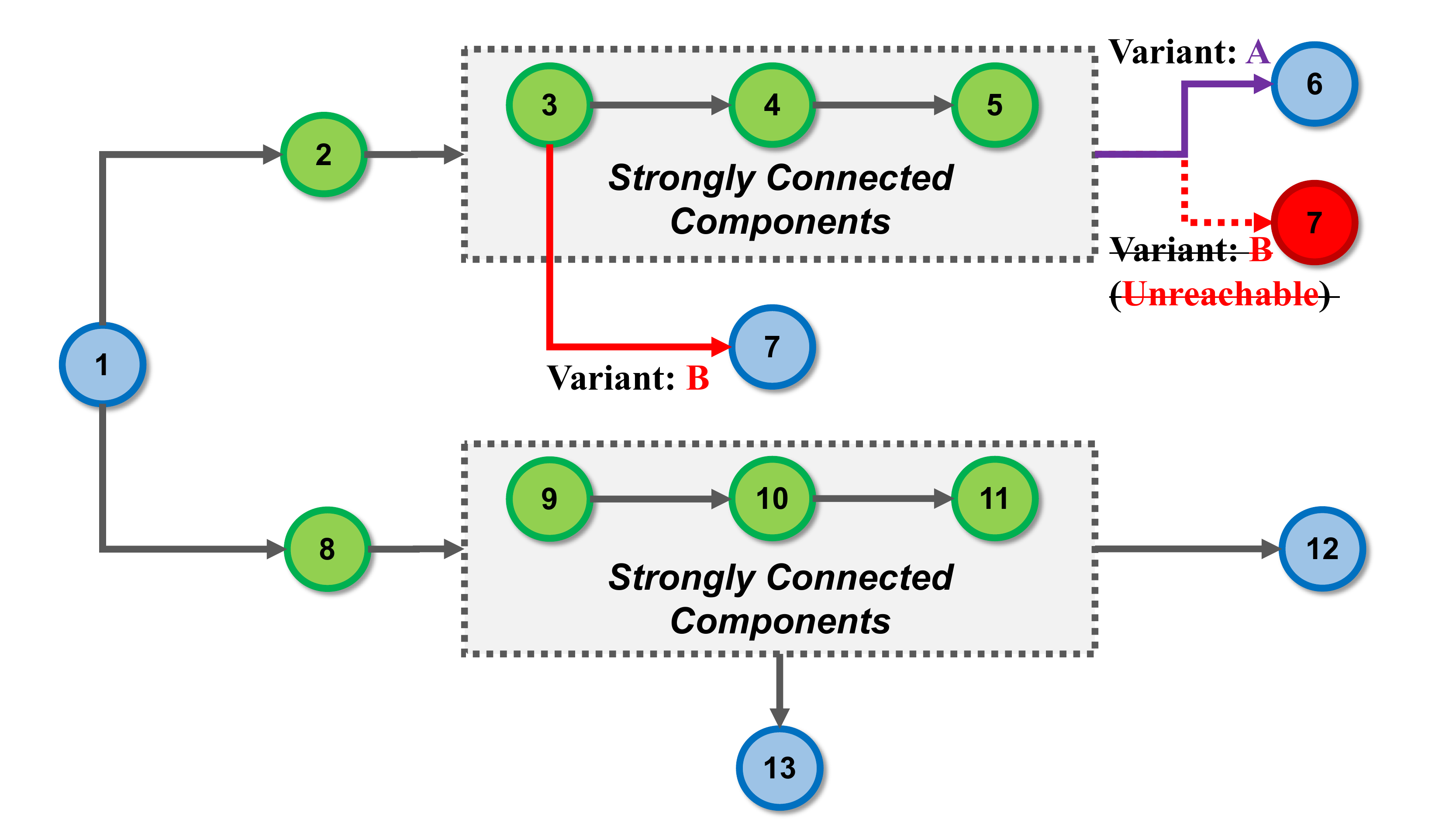}
	\caption{The final spanning tree generated for Figure~\ref{fig:control-flow-graph}.}
    \label{fig:spanning-tree}
\end{subfigure}
\caption{An example of the SwitchInt() control-flow and the final-generated spanning tree by using modified Tarjan algorithm. Variant A and variant B are the mutex variants towards the same enumeration and the terminator of node 3 is SwitchInt(). The final-generated spanning tree prunes the unreachable branches and establishes the independent branch for different variants while not affecting other kinds of terminators.}
\label{fig:tree}
\end{figure*}

Our approach leverages a modified Tarjan algorithm~\cite{10.1145/800061.808753} to remove the redundant paths. The details of the main algorithm can be found in Algorithm~\ref{algo:path}. The algorithm firstly uses a modified Trajan algorithm to generate a spanning tree of the graph (line 1-6) and then traverses this spanning tree to access all valuable paths (line 7). The traditional Tarjan algorithm is used to decompose strongly connected components (SCC) of the graph and removes the cycle succinctly (line 2). Therefore, SafeDrop can shrink the strongly connected components into points and generate a spanning tree of this graph (line 3). Ideally, the spanning tree enumerates all valuable paths without repeated traversal over cycles. However, such a coarse-grained approach may lead to incorrect analysis results due to some particular Rust statements. For example, Rust introduces a critical design towards the enumeration types that makes the traditional Tarjan algorithm less accurate, thus we should make some modifications. Figure~\ref{fig:tree} demonstrates such an exceptional case and our corresponding solution. 

To elaborate, Figure~\ref{fig:tree} employs a specific kind of terminator \texttt{SwitchInt()}, which is the culprit. The parameter of \texttt{SwitchInt()} is an enumeration, and the variants of the enumeration determine the control-flow \textit{e.g.,} node 3 in Figure~\ref{fig:control-flow-graph}. After shrinking SCCs into points, the branches connected with \texttt{SwitchInt()} in Figure~\ref{fig:control-flow-graph} are mutually exclusive, as variant A and variant B are mutex. SafeDrop therefore pinpoints the successive $k$ \texttt{SwitchInt()} towards the same enumeration parameters which contains $n$ possible variants (line 5), and then enumerate each variant once for all terminators to resolve mutex conflict (line 4-6). It prunes the unreachable paths and constructs independent paths for different variants (upper branch of node 1 in Figure~\ref{fig:spanning-tree}) without affecting other terminators (inferior branch of node 1 in Figure~\ref{fig:spanning-tree}). This method finally reduces the time complexity from $O(n^k)$ to $O(n)$ in this issue.

The traversal over the spanning tree will extract all valuable paths for SafeDrop. Since SafeDrop achieves a meet-over-paths scheme, it distributes the independent traversal state for each path. It backs up the state when meeting multi-branch nodes, and the state will be recovered when returning from the recursion (line 18). This operation promises the consistency of the traversal state when facing branches. Although we leverage a modified Tarjan algorithm to remove the redundant path, the path explosion is not fully inevitable. The unsettled path explosion is mainly generated from the nested conditional statements rather than cycles, so we set a large threshold value for it. If the counter exceeds the upper limit, SafeDrop will perform the traditional semi-lattice scheme and this is a trade-off between the precision and the speed.

\subsection{Alias Analysis}
SafeDrop performs alias analysis for each path and establishes the alias sets for each program point (line 11). In this subsection, we first discuss the basic rules for alias analysis and then present how we perform inter-procedural alias analysis.

\lstset{language=Rust,}
\begin{figure}[t]
\begin{lstlisting}[language=C, caption = The type filter in alias analysis. The composite types with an asterisk will be filtered if and only if all their fields are filtered in a recursive manner and these types should implement Copy trait at the same time.\label{list:filter}]

Value := Type::Bool         :  The primitive boolean type.
    | := Type::Char         :  The primitive character type.
    | := Type::Int          :  The primitive signed-integer type.
    | := Type::UInt         :  The primitive unsigned-integer type.
    | := Type::Float        :  The primitive floating-point type.
    | := Type::Array       *:  The homogeneous product types.
    | := Type::Structure   *:  The named product types.
    | := Type::Tuple       *:  The anonymous product types.
    | := Type::Enumeration *:  The tagged union types.
    | := Type::Union       *:  The untagged union types.
\end{lstlisting}
\end{figure}

\lstset{language=Rust,}
\begin{figure}[t]
\begin{lstlisting}[language=C, caption = The criteria of alias relationship and the matching operations to RValue in assignments. SafeDrop adopts a zero-overhead design for Move assignment because it will transfer the ownership of RValue as well as the alias relationship. SafeDrop constructs a map towards variables (fields) before analysis and transfers the map from RValue to LValue when meeting Move assignments.\label{list:alias}]

LValue := Use::Copy(RValue)      =>  e.g. _2 = _1         Use Copy Trait
    |  := Use::Move(RValue)   *  =>  e.g. _2 = move _1    Use Drop Trait
    |  := Cast::Copy(RValue)     =>  e.g. _2 = _1 as i32  Cast Types
    |  := Cast::Move(RValue)  *  =>  e.g. _2 = move _1 as i32
    |  := Ref(RValue)            =>  e.g. _2 = &mut _1    Create Reference
    |  := AddressOf(RValue)      =>  e.g. _2 = *mut _1    Create Raw Pointer
\end{lstlisting}
\end{figure}

\subsubsection{\textbf{Basic Rules for Alias Analysis.}} Not all aliases are our interest. Since memory deallocation bugs are generally triggered by \texttt{Drop} trait, SafeDrop introduces a type filter to skip the analysis towards the \texttt{Copy} trait (stack-only) variables and omit to analyze the statements with the filtered types. Listing~\ref{list:filter} extracts irrelevant types that can be ignored in this procedure. The Rust primitives which are \texttt{Copy} trait would be simply filtered \textit{i.e.} \texttt{i32}. As an exception, it deletes slice, reference and raw pointer types in the filter due to their precise transmission of alias relationship, even if they are \texttt{Copy} trait. SafeDrop expands the filter to the composite types and checks each field recursively \textit{i.e.} \texttt{Box<T>}. For each composite type, it will be filtered as long as all its components are checked as filtered and it should implement \texttt{Copy} trait at the same time.

Listing~\ref{list:alias} summarizes 5 kinds of statements that contribute to alias relationships. These statements are assignments with the different operations to \texttt{RValue}. After applying the type filter to the variables, the matching statements would establish the alias relationship for \texttt{RValue} and \texttt{LValue}. In SafeDrop, we use union-find disjoint sets to store the alias relationship. Set \texttt{A} and set \texttt{B} will be merged if one element in set \texttt{A} has an alias relationship with the element in set \texttt{B}. We provide an example of this alias establishment. If both \texttt{LValue} and \texttt{RValue} in an assignment are not filtered and the assignment matches the form of \texttt{LValue:= Ref(RValue)}, SafeDrop would establish the alias relationship and merge the sets.

SafeDrop is also a field-sensitive approach because the field-insensitive strategy will incur a bunch of false positives into SafeDrop. Since it applies the union-find disjoint sets to represent the alias relationship, it is far less accurate and is prone to merge irrelevant sets if we consider the whole composite-type variable as a unit instead of splitting its fields. Therefore, SafeDrop identifies whether \texttt{LValue} and \texttt{RValue} in one assignment is a field of a composite-type variable and establishes the alias set for the field independently. This method is also valid for parameter passing and value returning in the inter-procedural alias analysis as follows.

\subsubsection{\textbf{Inter-Procedural Alias Analysis}} The basic rules for alias analysis is an intro-procedural approach. It analyzes each statement flow-sensitively and stores the alias relationships as disjoint sets for each program point. However, this analysis could be unsound due to function calls. If a path contains function calls, we should perform inter-procedural analysis to obtain the alias relationship between parameters and return value.

\textbf{Callee Analysis.} The call chain of a program is embedded in the terminator of each basic block, containing parameters, return value, and the internal callee ID. As the arrival of a function call, SafeDrop invokes a query to ask the compiler for the final-optimized MIR of the callee through its internal ID. SafeDrop then traverses this MIR and establishes the alias sets for its variables. The alias analysis of callee finally returns a result containing the alias relationship between arguments and return value to its caller.

\textbf{Recursion Refinement.} SafeDrop adopts a fixed-point iteration method to solve the recursive invocation problem. It maintains a call stack to ensure the rule that each function ID can only appear once in this stack. As a function first arriving and pushing its ID into the call stack, SafeDrop will set default alias relationship (false) between return value and arguments, and abort to prevent infinite recursion. Considering the recursive invocation would generally exit and the alias relationship may change at the same time, the analyzer should re-execute SafeDrop to the function in the stack with the updated alias results until the final alias relationship is reaching the fixed-point.

\textbf{Cache Refinement.} Since SafeDrop is a meet-over-paths approach, the alias result between arguments and return value is the union of all valuable paths (line 15). Namely, one argument has the alias relationship with the return value as long as it has such a relationship at least one path. SafeDrop only needs to analyze once towards the given function. After traversing the function for the first time, the analysis result will be cached into the hash table (line 9), and the analyzer can get the result directly when encountering this function again.

\subsection{Rules for Invalid Drop Detection}
The safety of one deallocation generally depends on the alias sets. For each path, SafeDrop detects each memory deallocation flow-sensitively to confirm whether it would incur memory-safety issues (line 12). If a memory deallocation bug is detected, SafeDrop will record the kind of this issue as well as the related source code, then merge the result at the end of the path (line 15).

SafeDrop maintains a taint set to record the deallocated buffers, as well as returned dangling pointers. It adds the dropping variable into the taint set and marks it as the taint source when finding \texttt{Drop()} in the terminator. As for a composite-type variable, SafeDrop will add each drop-trait field into the taint set rather than the entire variable. The taint source propagates in the alias set and pollutes other aliases. SafeDrop then iteratively checks the alias relationship between the using variable and all elements in the taint set at each program point. As an exception, we insert the variable constructed from \texttt{uninitialized()} to taint set as the time of declaration, and remove this variable if it is initialized later.

According to the typical patterns in section 3, we summarize four rules for the memory-safety issues caused by invalid memory deallocation, and all of these rules would be applied for normal execution and exception handling path as follows.

\begin{itemize}
\item \textbf{Use After Free:} The taint set contains the alias of the using variable in a statement or passing this variable to a function.
\item \textbf{Double Free:} The taint set contains the alias of the dropping variable in a \texttt{Drop()} terminator.
\item \textbf{Invalid Memory Access:} The taint set contains an uninitialized variable at the time of using and dropping.
\item \textbf{Dangling Pointer:} The taint set contains the alias of the returned pointer. Although it loses the using context, this pointer is actually buggy and highly unsafe to use.
\end{itemize}

SafeDrop employs independent flags for these rules. The true flag means the existence of the specific memory-safety issue inside this function. These flags are cached into the hash table as part of the analysis result after the inter-procedural analysis (line 9). For each path, these flags will be merged as well as the alias relationship between arguments and return value, when SafeDrop completes the traversing over the selected path (line 15).

\subsection{Corner Case Handling}

\begin{figure}[t!b]
\includegraphics[width=0.45\textwidth]{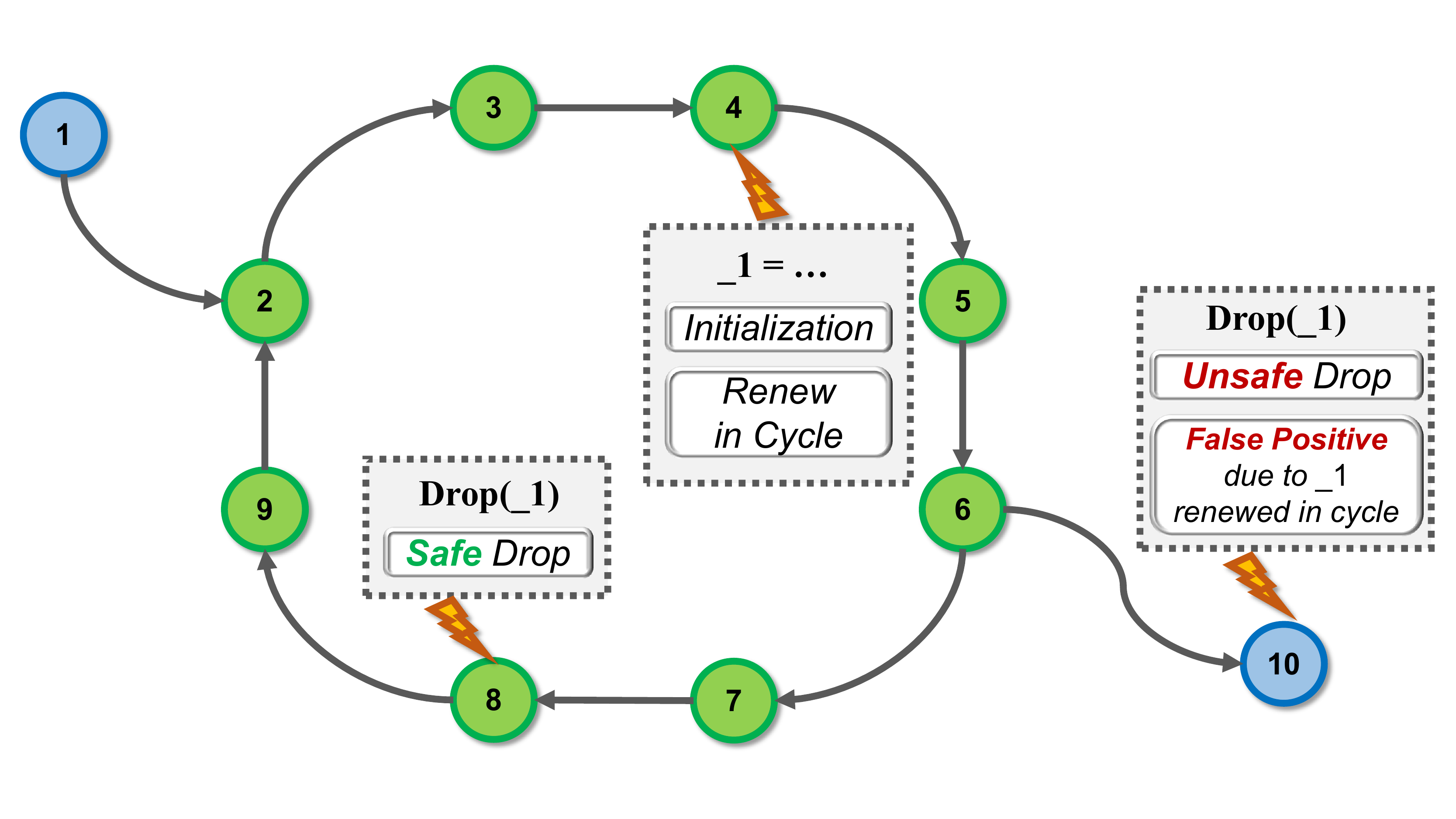}
\caption{A corner case example of renewing after dropping in cycle. The refinement of such issue regards the dealloction in node 10 as a safe drop in the path 1-(2-3-4-5-6-7-8-9)-10.}
\label{fig:drop-renew}
\end{figure}

The modified Tarjan algorithm generates a spanning tree to extract valuable path that removes the redundant traversal over cycles. However, there exists one kind of corner case that introduces a bunch of false positives and we use Figure~\ref{fig:drop-renew} as an example. In Figure~\ref{fig:drop-renew}, \texttt{\_1} is initialized and dropped in SCC (2-3-4-5-6-7-8-9). The deallocation of node 10 in the path 1-2-3-4-5-6-10 is safe, and is regarded as an invalid drop in path 1-(2-3-4-5-6-7-8-9)-10. The second path is indeed a false positive due to \texttt{\_1} is renewed in node 4, but we omit traversing twice from node 2 to node 6. Thus we make the refinement that SafeDrop records the definition block of the variables and verifies whether the branch targeted to the repeated drop towards the same variable is between the definition block and deallocation block in SCC. The correctness of the refinement is based on the monotonic property of the alias relationship merging in alias analysis.

\section{Evaluation}

\subsection{Implementation}
We implement our approach as a query in Rust MIR considering that MIR desugars most of Rust's surface representation, leaving a simpler form for type-checking, translation, and safety analysis. This query will be invoked after running all optimization passes in MIR and can be triggered as needed during compilation to check the whole functions in each crate. SafeDrop is integrated into the Rust compiler v1.52 and can be used by the command line tools such as \texttt{rustc} and \texttt{cargo}.

\subsection{Experimental Setup}
This section presents our experimental setting for evaluating the effectiveness, applicability and efficiency of SafeDrop. Specifically, we focus on the questions as follows.

\textbf{RQ1:} Can SafeDrop recall existing CVEs of such memory deallocation bugs? How many false positives incurred?

\textbf{RQ2:} Is SafeDrop applicable for helping users to alert memory deallocation bugs? Is the number of warning manageable?

\textbf{RQ3:} How much overhead does SafeDrop introduce and slow down the compilation process?

To answer RQ1, we collect a dataset with all related CVEs for evaluation~\cite{xu2020memory}, including 2 use-after-free issues, 3 double-free issues, and 4 invalid memory access issues. These 9 CVEs are from 8 different Rust crates.

To answer RQ2, we evaluate 24 real-world Rust crates from GitHub and apply SafeDrop to find invalid memory deallocation issues.

To answer RQ3, we measure the compilation time of the CVE crates with SafeDrop and calculate the overhead compared with the original compiler.

Our experiments are done on a 2.00GHZ Intel processor with the 18.04.1-Ubuntu operating system (Linux kernel version 5.4). The time measured in the experiment is the average of 3 runs. 
\subsection{Results and Analysis}

\subsubsection{\textbf{Effectiveness.}}

\begin{table*}[t!b]\small
  \caption{Experimental results of the vulnerabilities reported in CVEs. These CVEs are containing four kinds of bugs including use-after-free (UAF), double-free (DF), dangling pointer (DP), and invalid memory access (IMA). The results have been collected and classified with manually checking. The bugs reported are classified as True Positive/False Positive (TP/FP) in the table (using dash to simplify the representation of 0/0), and we also count the number of the methods and the code lines to represent the crate scale.}
  \label{CVE}
  \centering
  \begin{tabular}{lcccccc>{\columncolor{mygray}}cccr}
    \toprule[1pt]
    \multicolumn{1}{c}{\multirow{2}{*}{\textbf{Crate}}} 
    & \multirow{2}{*}{\textbf{CVE}} 
    & \multirow{2}{*}{\textbf{CVE-Type}} 
    & \multicolumn{5}{c}{\textbf{SafeDrop Report (TP/FP)}} 
    &\multirow{2}{*}{\textbf{Recall}} 
    & \multirow{2}{*}{\textbf{Methods}} 
    & \multirow{2}{*}{\textbf{Lines}}\\
    \cmidrule(r){4-8}
     & & & UAF & DF &  DP & IMA & Total &  &  & \\
    \midrule[0.6pt]
    
    isahc                      & 2019-16140 & UAF & - & 0/1  & 1/0 & -   & 1/1    & 100\% & 89   & 1304\\ 
    open-ssl                    & 2018-20997 & UAF & 1/2 & -  & 0/1 & -   & 1/3    & 100\% & 1188 & 20764\\
    linea                      & 2019-16880 & DF  & - & 1/0  & - & 10/0 & 11/0    & 100\% & 1810 & 24317\\
  
    ordnung                    & 2020-35891 & DF  & 0/1 & -  & 3/0 & -   & 3/1    & 100\% & 145  & 2546\\ 
    \multirow{2}{*}{smallvec}  & 2018-20991 & DF  & \multirow{2}{*}{-}   & \multirow{2}{*}{-}  &  \multirow{2}{*}{1/2} & \multirow{2}{*}{1/0}  &    & \multirow{2}{*}{100\%} & \multirow{2}{*}{187}  & \multirow{2}{*}{2297}\\
                                        & 2019-15551  & DF  &     &    &     &     &  \multirow{-2}*{2/2}  &      &    \\
                                        
    crossbeam                 & 2018-20996  & IMA  & - & 0/1 & - & 2/0   & 2/1    & 100\% & 221  & 4184\\
    generator                 & 2019-16144  & IMA  & - & - & - & 1/0   & 1/0    & 100\% & 158  & 2608\\
    linkedhashmap             & 2020-25573  & IMA  & - & - & - & 1/0   & 1/0    & 100\% & 137   & 1974\\
    \bottomrule[1pt]
  \end{tabular}
\end{table*}

We perform SafeDrop towards the reported CVEs to testify whether our approach is effective to locate the invalid memory deallocation bugs. Table~\ref{CVE} lists a series of related CVEs containing invalid memory deallocation issues. For each crate, we collect the analysis results and classify warnings into true positives and false positives. SafeDrop can recall all related vulnerabilities of the reported CVEs listed in Table~\ref{CVE} that represents the effectiveness of SafeDrop.

The classification of the warnings depends on the rules for invalid drop detection introduced in Section 4. We manually check the output to enhance the accuracy of this classification. The count of true positives and false positives are mostly based on these rules, but some memory-safety issues in callee will intensify its caller. This scenario is common for dangling pointers \textit{e.g.,} if the callee returns a dangling pointer, using or dropping this pointer in its caller will incur use-after-free or double-free bugs. Therefore, the classification is not confined to the CVE type.

The effectiveness evaluation does not calculate the false positive rate due to SafeDrop only detecting the invalid deallocation bugs incurred by OBRM that has a relatively low cardinal. The result demonstrates the quantity of false positives is limited in SafeDrop. Moreover, the program with a large scale does not tend to trigger more false positives as shown in Table~\ref{CVE}. The code lines of crate \texttt{open-ssl} is ninefold larger than crate \texttt{smallvec}. However, the number of false positives differs in 1 only. Therefore, the quantity of false positives is manageable, and the programmer can easily locate the innermost function and make corrections based on warning info. 

\subsubsection{\textbf{Applicability}} 

\begin{table}[]\small
  \caption{Experimental results of the applicability. We select a series of crates with the usage of unsafe constructor and perform SafeDrop to evaluate these crates. The collection and classification is same as Table~\ref{CVE}. These new issues previously unknown are from 8 different crates.}
  \label{new_bugs}
  \centering
\begin{tabular}{lcccc>{\columncolor{mygray}}r}
\toprule[1pt]
\multicolumn{1}{c}{\multirow{2}{*}{\textbf{Crate}}} & \multicolumn{5}{c}{\textbf{SafeDrop Report (TP/FP)}}  \\
\cmidrule(r){2-6}
                      & UAF & DF &  DP & IMA & Total   \\ \midrule[0.6pt]
                      
wasm-gb               & -  & -  &  20/0  & -   & 20/0    \\ 
bzip2                 & -  & -  &  2/1    & -   & 2/1       \\ 
rust-poker            & -  & 1/0  &  -    & -   & 1/0       \\ 
wasm-integration      & -  & -  &  2/0    & -   & 2/0      \\ 
array                 & -  & -  &  1/0    & -   & 1/0   \\ 
teardown-tree         & -  & 0/2  &  1/0    & -   & 1/2  \\ 
apres-bindings        & -  & -  &  14/0  & -   & 14/0  \\ 
rust-workshop         & -  & -  &  2/0  & -     & 2/0  \\ 
\bottomrule[1pt]
\end{tabular}
\end{table}

\begin{figure}
\centering
\includegraphics[width=0.45\textwidth]{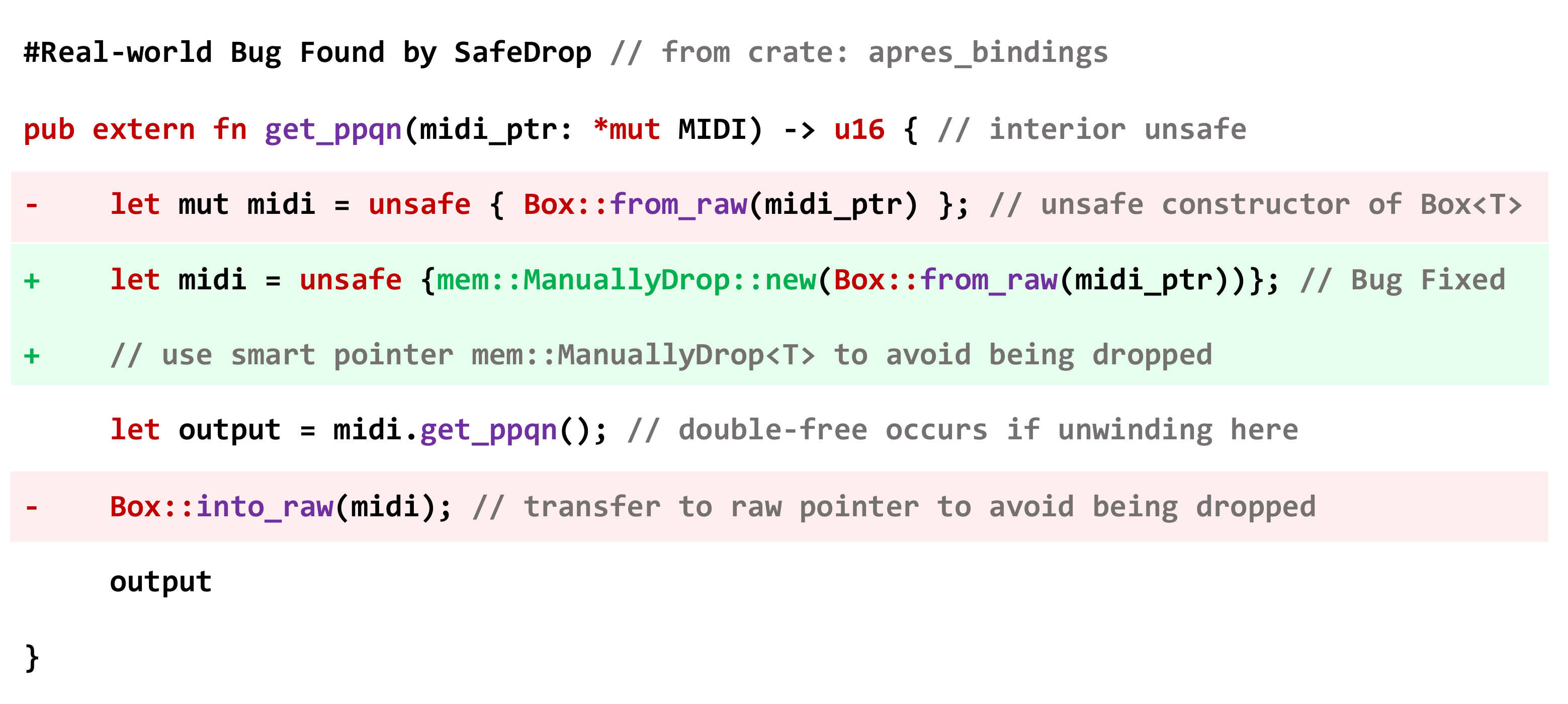}
\caption{An example of the new bug found by SafeDrop. This example accords with the definition of dropping buffers in use introduced in Section 3. It is an invalid drop of exception handling and matches pattern 2 in List~\ref{list:pattern}.}
\label{code:new_bugs}
\end{figure}

We perform SafeDrop to detect the invalid memory deallocation bugs in some real-world Rust crates. The prototype of SafeDrop is a static analysis pass of the compiler. Ideally, it can print warnings to be applicable for users to find the potential bugs with false positives as few as possible. Therefore, developers can manually check the buggy snippets, then make some corrections, and finally re-compile the whole crate through SafeDrop to verify whether the warning is eliminated.

Table~\ref{new_bugs} shows the applicability of SafeDrop and lists the finding issues previously unknown caused by invalid memory deallocation. We collect 24 crates from Github with the criteria of using unsafe constructor. After running SafeDrop to these crates, there are 8 crates having such issues as listed in Table~\ref{new_bugs}. For each crate, we collect the analysis results and classify them like the effectiveness evaluation. The results demonstrate that the quantity of the reported false positives is still at a low-level and invalid memory deallocation issues are common in unsafe Rust. In particular, SafeDrop introduces 0 false positive in $75\%$ of these crates and produces no more than 2 false positives in other crates. Therefore, SafeDrop has good applicability for developers to find and locate such bugs during compilation time.

Figure~\ref{code:new_bugs} shows an example of new bugs found by SafeDrop and this bug is confirmed by the developer in Github. In the buggy snippet, there is a vulnerable program point \texttt{midi.get\_ppqn()} that is between \texttt{Box::from\_raw()} and \texttt{Box::into\_raw()}. Unwinding at this point will incur a double-free bug. This bug is similar to the first motivating example and matches pattern 2 in Section 3. In fact, the issues in Figure~\ref{code:new_bugs} are generally be triggered in the unwind path, because glibc fasttop~\cite{glibc} can detect double free when running programs after compilation. That motivates the developers to find the buggy code and makes such normal execution bugs rare to find. However, the exception handling bugs are hard to make corrections and we advise using \texttt{ManuallyDrop<T>} to this scenario as shown in Figure~\ref{code:new_bugs}.


\subsubsection{\textbf{Efficiency.}} 

\begin{figure}[t!b]
\includegraphics[width=0.47\textwidth]{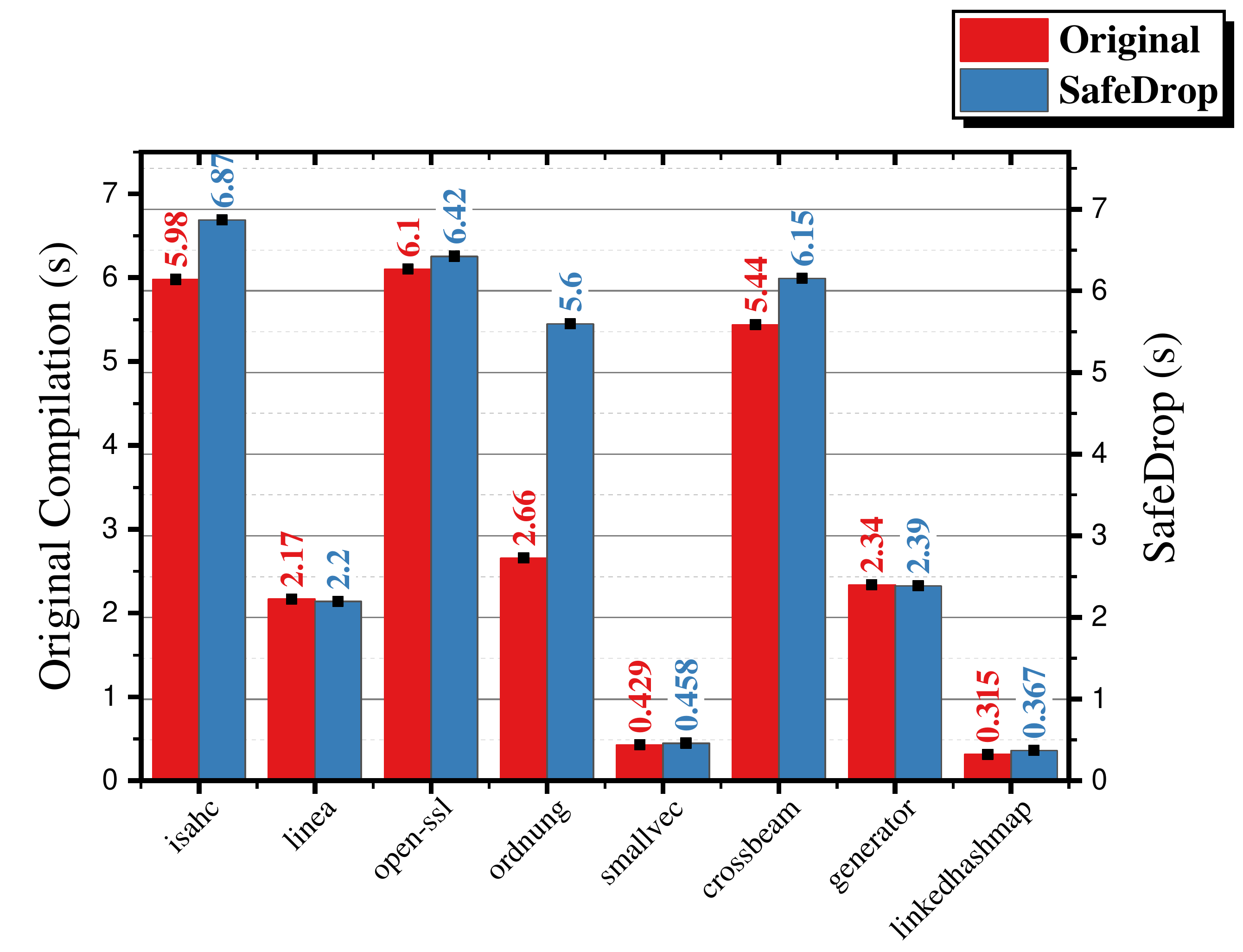}
\caption{Experimental results of the efficiency. We compare the compilation time between the original compiler and SafeDrop towards all CVE crates. The overhead will not exceed $2\times$ to the original compiling time in general. Crate \texttt{ordnung} has an apparently large overhead due to amounts of successive \texttt{SwitchInt()} nodes with the changing variants in the control-flow graph. It will produce more branches in the spanning tree that enlarges the analyzing time.}
\label{fig:efficiency}
\end{figure}

SafeDrop is a query integrated into the Rust compiler. Ideally, it should not drag much compilation time. We evaluate the overhead in comparison with the original compiler.

The compilation time is composed of the evaluated crates as well as their dependencies crates. Therefore, it is equivalent to involve the dependencies to represent the whole compilation time without affecting the overhead ratio. Figure~\ref{fig:efficiency} shows the efficiency evaluation results of 8 CVE crates. The overhead ranges from 1.2\% to 110.7\% and the average overhead is in the region (0\%,120\%]. We can conclude that the overhead introduced by SafeDrop would not exceed $2\times$ to the original compiling time in general and SafeDrop emerges an excellent efficiency towards the different scales of the crates. Moreover, the results demonstrate the irrelevance between the scale and the overhead ratio. As shown in Figure~\ref{fig:efficiency}, the overhead ratio of crate \texttt{open-ssl} and \texttt{smallvec} differs in 1.6\%, which are 5.2\% and 6.8\% respectively. The scale of \texttt{open-ssl} is ninefold larger along with a lower overhead ratio that proves our conclusion.

\subsection{Discussion} 
\subsubsection{\textbf{False Positives.}} The inter-procedural alias analysis heavily relies on the query \texttt{optimized\_mir} to obtain MIR of the callees. However, SafeDrop cannot capture some MIR such as specific trait implementations due to the internal design of the compiler. After running the type filter, our approach then establishes the alias relationships between unfiltered arguments and return value. We adopt this operation to pursue the soundness of alias recognition that would lead to some false positives rather than losing alias relationships. Since these cases are not pervasive and the type filter discerns most of the evident errors, the false positives are manageable in SafeDrop. In particular, we remove the alias established from \texttt{Clone} trait \cite{klabnik2019rust}, because the return value of the function \texttt{clone()} is a deep copy to the original variable.

\subsubsection{\textbf{False Negatives.}} Rust adopts several tricky designs to some inline functions that may incur false negatives. The query can successfully get MIR towards these functions but containing empty information. Thus SafeDrop will lose some alias relationships. Additionally, SafeDrop does not provide the support for primitive array type, closure, raw pointer offset, and function pointer that will introduce false negatives.

\section{Related Work}
In this section, we will introduce some related work and compare them with SafeDrop to discuss the significance of our approach.

In the past few years, the existing work towards Rust programming language mainly focus on formal verification~\cite{crust,RustTypesV} and unsafe code use~\cite{UnsafeComponents} through dynamic~\cite{fuzz_rust} or static approaches. RustBelt~\cite{Rustbelt} is the first formal tool for verifying the safe encapsulation of unsafe code in Rust. XRust~\cite{XRust} is a heap allocator that isolates the memory of unsafe Rust and prevents any cross-region memory corruption. RustBelt, XRust and other researchs~\cite{Rustbelt_relax} based on them separate processing safe Rust and unsafe Rust to enhance memory safety. However, SafeDrop merges this separation in MIR and checks the validity of each deallocation to enhance safety.

Rust Miri~\cite{miri} is an experimental interpreter for Rust MIR, which can run binaries, test suites of cargo projects and detect some undefined behaviors. The functionality of Miri is wider than SafeDrop and Miri supports the checking of out-of-bounds memory accesses, insufficient aligned memory accesses and violations of some basic type invariants in addition. However, Miri is a dynamic analysis approach and cannot track all valuable paths of a program that is not capable of unwinding analysis and library analysis. On the contrary, SafeDrop is a static approach that can analyze all valuable path of each function in compile-time .

There are numerous work proposed to detect memory-safety issues~\cite{MCWeakly,olivo2015static,DangSan,Spatio}. Most of these work are based on the specified programming language, we list several generic tools as follows. Valgrind (Memcheck)~\cite{Valgrind} is an instrumentation framework that can automatically detect many memory management bugs. AddressSanitizer~\cite{AddressSanitizer} is another efficient memory error detector based on LLVM. Since Valgrind and AddressSanitizer are dynamic analysis approaches like Rust Miri, the analysis of SafeDrop is in the compile-time and will not affect the running speed of the program. 

\section{Conclusion}
In this paper, we proposed SafeDrop, a novel compiler-integrated path-sensitive data-flow approach to detect memory deallocation violations in Rust. We implemented SafeDrop, applied it to the Rust compiler, and conducted a thorough evaluation with existing Rust CVEs and real-world Rust crates to show its value for memory deallocation bugs detection. Our results demonstrated the effectiveness, applicability and efficiency of SafeDrop. Moreover, our application of SafeDrop identified 8 Rust crates involved with invalid memory deallocation issues previously unknown. We believe the method of SafeDrop can be used for the specific kind of security detections, including use-after-free, double-free, dangling pointer, and invalid memory access caused by automatic memory deallocation in other RAII systems.

\bibliographystyle{ACM-Reference-Format}
\bibliography{safedrop}

\end{document}